\documentclass[prc,showpacs,preprintnumbers,showkeys,superscriptaddress]{revtex4}

\usepackage{enumerate} 
\usepackage{amsmath}  
\usepackage{amsthm}  
\usepackage{amsfonts}  
\usepackage{amssymb} 
\usepackage{amsxtra}
\usepackage{amssymb}
\usepackage{delarray} 
\usepackage{graphicx}
\usepackage{dcolumn}  
\usepackage{epsfig}  
\usepackage{float}
\usepackage{hhline}
\restylefloat{figure}

\newcommand{\Sigsn}{\Sigma_{{\mathrm s},n} }

\newcommand{\Sigon}{\Sigma_{{0},n} }
\newcommand{\Sigsp}{\Sigma_{{\mathrm s},p} }

\newcommand{\Sigop}{\Sigma_{{0},p} }
\newcommand{\Sigsi}{\Sigma_{{\mathrm s},i} }
\newcommand{\Sigvi}{\Sigma_{{\mathrm v},i} }
\newcommand{\Sigoi}{\Sigma_{{0},i} }

\DeclareMathOperator{\Tr}{Tr}

\begin{document}

\title{Nuclear matter in the crust of neutron stars derived from realistic NN
interactions}
\author{P. G\"{o}gelein}
\affiliation{Institut f\"{u}r Theoretische Physik,Universit\"{a}t T\"{u}bingen, 
D-72076 T\"{u}bingen, Germany}
\author{E.N.E. van Dalen}
\affiliation{Departament d'Estructura i Constituents de la Mat\`eria,
Universitat de Barcelona, Diagonal 647, E-08028 Barcelona, Spain}
\affiliation{Institut f\"{u}r Theoretische Physik,Universit\"{a}t T\"{u}bingen, 
D-72076 T\"{u}bingen, Germany}
\author{C. Fuchs}
\affiliation{Institut f\"{u}r Theoretische Physik,Universit\"{a}t T\"{u}bingen, 
D-72076 T\"{u}bingen, Germany}
\author{H. M\"{u}ther}
\affiliation{Institut f\"{u}r Theoretische Physik,Universit\"{a}t T\"{u}bingen, 
D-72076 T\"{u}bingen, Germany}



\begin{abstract}
Properties of inhomogeneous nuclear matter are evaluated within a relativistic
mean field approximation using density dependent coupling constants. A
parameterization for these coupling constants is presented, which reproduces the
properties of the nucleon self-energy obtained in Dirac Brueckner Hartree 
Fock calculations of asymmetric nuclear matter but also provides a good
description for bulk properties of finite nuclei. The inhomogeneous
infinite matter is described in terms of cubic Wigner-Seitz cells, which allows
for a microscopic description of the structures in the so-called ``pasta-phase''
of nuclear configurations and provides a smooth transition to the limit of
homogeneous matter. The effects of pairing properties and finite temperature are
considered. A comparison is made to corresponding results employing the
phenomenological Skyrme Hartree-Fock approach and the consequences for the
Thomas-Fermi approximation are discussed. 
\end{abstract}


\maketitle

\section{Introduction}

One of the main goals of microscopic nuclear structure calculations, which are
based on realistic models for the nucleon-nucleon (NN) interactions, is to
obtain a reliable prediction for the equation of state of nuclear matter under the
extreme conditions, which one has to consider for astrophysical scenarios like a
supernova or objects like a neutron star. For that purpose one considers NN
interactions, which are adjusted to describe the properties of two nucleons in
the vacuum, i.e. the NN phase shifts, and tries to develop a many-body theory
which yields a good description for the bulk properties of ``normal'' nuclear
matter, the saturation point of symmetric nuclear matter and properties of
finite nuclei. A theory, which is able to link such a realistic NN interaction
to the properties of nuclear matter at normal densities, should provide reliable
results for the properties of nuclear matter at higher densities, temperatures
and isospin-asymmetries, as they occur in the astrophysical objects mentioned
above\cite{bal1}. 

A basic problem of such nuclear structure investigations is the existence of
strong tensor short-range components in such a realistic NN interaction, which
makes it necessary to account for corresponding correlations in the nuclear
wave-function. In fact, simple Hartree-Fock or mean-field calculations using
such realistic NN interactions yield unbound nuclei\cite{her1}. Rather powerful
many-body techniques have been developed to account for correlations beyond the
mean-field approximation and using rather sophisticated Monte-Carlo techniques
one is able to derive the properties of light nuclei from such realistic NN
interactions\cite{pieper02}. In order to obtain a good agreement with the
experimental data, however, one has to introduce a three-body force.

An additional three-body force is also required, if one wants to reproduce the
saturation point of symmetric nuclear matter within a non-relativistic many-body
approach based on realistic NN interactions. On the other hand, however, it is
known already for more than 15 years that the consideration of relativistic
features, as it is done e.g. in the Dirac  Brueckner  Hartree  Fock (DBHF)
approximation yields results for the saturation point of symmetric nuclear
matter, which are very close to the empirical
data\cite{anast,brock,malf1,weigel}. 

This success is based on the relativistic structure of the NN interaction, a
feature which is also present in the phenomenology of the Walecka
model\cite{serot}.    The NN interaction in this model is described in terms of
the exchange of a scalar meson, $\sigma$, and a vector meson $\omega$.
Calculating the nucleon self-energy, $\Sigma$, from such a meson exchange model
within a mean-field (Hartree) approximation, one finds that the $\omega$ 
exchange yields a component $\Sigma^0$, which transforms under a Lorentz
transformation like the time-like component of a vector, while the scalar meson
exchange yields a contribution $\Sigma^s$, which transforms like a scalar.
Inserting this self energy into the Dirac equation for a nucleon in the medium
of nuclear matter leads to single-particle energies,  which are as small as the
empirical value of -50 MeV. This small binding effect, however, results from a
strong cancellation between the repulsive $\Sigma^0$ and the attractive
$\Sigma^s$ component. 
As has recently been shown in \cite{plohl06} the appearence of these two 
large and cancelling scalar and vector fields in matter is not only a 
consequence of Dirac phenomenology but has a deeper foundation in the 
structure of the nucleon-nucleon force, i.e. it is intimitely 
connected to the spin-orbit force in vacuum NN scattering.
The scalar component $\Sigma^s$, which is conveniently
described in terms of an effective Dirac mass, leads to a significant enhancement
of the small component of the Dirac spinors in the nuclear medium. It is this
density-dependence of the Dirac spinors, which is responsible for the
fine-tuning in the nuclear structure calculation, which is necessary to obtain
the empirical saturation point. Non-relativistic studies, which cannot account
for this feature, may include this effect in terms of a three-body force.
Moreover, the excitation of anti-nucleons, automatically included 
in the relativistic formalism, gives rise to a class of three-body forces 
(Z-graphs) which have to added in the non-relativistic formalism. Doing so, 
the saturation points from non-relativistic Brueckner calculations are 
shifted towards their relativistic counterparts \cite{Li:2006gr}. 

A neutron star, however, cannot completely be described in terms of homogeneous
nuclear matter in $\beta$-equilibrium at various densities. The crust of such a
neutron star in particular is a challenge for theoretical nuclear
structure physics as it incorporates the transition from isolated nuclei via a
crystal-like structure of quasi-nuclei embedded in a sea of neutrons to phase of
homogeneous baryon matter. These intermediate structures have been described by
means of the Wigner-Seitz (WS) cell approximation employing the Thomas-Fermi
approximation based on a purely phenomenological energy-density functional. Such
calculations predicted a variety of geometrical structures: 
Spherical quasi-nuclei, which are favored at small
densities, merge with increasing density to strings, which then may cluster to
parallel plates and so on. These geometrical structures have been the origin for
the popular name of this phase: Pasta phase. 

A step towards a more microscopic study has been made by performing
self-consistent Hartree-Fock calculations based on effective nucleon-nucleon
interactions like the density-dependent Skyrme forces\cite{sk1,sk2}. Such
calculations have been performed  by Bonche and Vautherin\cite{bv81} and by  a
few other groups. These studies show that shell effects have a significant 
influence on details like the proton fraction of the baryonic matter in the 
inhomogeneous phase\cite{Montani:2004}. They also provide the basis for a
microscopic investigation of pairing phenomena, excitation modes and
response functions as well as the effects of finite temperature. 

Such self-consistent calculations for inhomogeneous matter have typically been
performed assuming a WS cell of spherical shape. This geometry, however, does
not support the description of triaxial structures, which are typical for the
pasta phase. Furthermore the transition to homogeneous matter cannot be described
in a satisfactory manner within such a spherical WS cell\cite{Montani:2004}.

Therefore cubic WS cells have recently been considered for Hartree-Fock
calculations using effective Hamiltonians like the Skyrme
forces\cite{Goegelein:2007}. These effective Hamiltonians, however, are of
purely phenomenological origin. They are adjusted to describe nuclear matter at
normal density and finite nuclei. Therefore one cannot expect any predictive
power of such calculations if one extends its region of application to higher
densities, large proton-neutron asymmetries and finite temperatures.

Therefore in the present work we will perform relativistic mean field
calculations employing a model for the Lagrangian, which is based on
microscopic  DBHF calculations using  realistic NN interactions. Attempts to
derive an effective Lagrange density by fitting density-dependent coupling
constants in such a way that a mean-field calculation reproduces details of a
DBHF calculation have been made before\cite{Schiller:2001,Hofmann:2001}. The
calculations presented here are based on DBHF calculations of van Dalen et
al.\cite{vanDalen:2007} for asymmetric nuclear matter. These results for the 
DBHF calculations for homogeneous asymmetric nuclear matter have successfully 
been used in bulk studies of neutron stars\cite{Klaehn:2006}. Therefore we have
tried to ensure that the effective mean field parameterization reproduces the
properties of the equation of state derived from this DBHF calculations at high densities
with good accuracy. Furthermore we introduce a small correction term, which
ensures that the mean-field calculations yield a good description of binding
energy and radii of finite nuclei as well. In this way we obtain an effective
field theory, which is based on a realistic model for the NN interaction in the
vacuum, reproduces the details of microscopic DBHF calculations for asymmetric
nuclear matter at high densities and yields good agreement with the empirical
data for the saturation point of asymmetric nuclear matter as well as bulk
properties of finite nuclei. Therefore the resulting parameterization should be a
good candidate to determine a reliable equation of state for baryonic matter
in a large interval of densities.

Using this Lagrangian we will present results of relativistic mean-field
calculations in Cartesian WS cells considering a range of densities in which the
transition from isolated nuclei to homogeneous matter occurs. Special attention
will be paid to the effects of finite temperature on the formation of the
geometrical structures. 

After this introduction section 2 contains a short review of the Density
Dependent Relativistic Mean Field (DDRMF) approach and its application to the
description of infinite matter as well as finite nuclear systems and baryonic
structures in a Cartesian WS cell. The parameterization of the density dependent
coupling constants and the fit to the DBHF results is described in section 3.
Results for the inhomogeneous structures, which are typical for the crust of 
neutron stars are presented in section 3. Special emphasis is made to explore
the effects of finite temperature and the differences as compared to studies
using a purely phenomenological Skyrme forces. 

\section{The DDRMF Approach}

The Density Dependent Relativistic Mean Field (DDRMF) approach 
is an effective Field theory of interacting mesons and nucleons. Following the
usual notation we consider scalar ($\sigma,\delta$) and vector mesons
($\omega,\rho$), which with respect to the isospin correspond to isoscalar
($\sigma,\omega$) and isovector ($\delta,\rho$), respectively. The Lagrangian
density consists of three parts: 
the free baryon Lagrangian density $\mathcal{L}_B$, 
the free meson Lagrangian density $\mathcal{L}_M$
and the interaction Lagrangian density $\mathcal{L}_{\text{int}}$:
\begin{equation}\label{Lag_dens}
	\mathcal{L} = \mathcal{L}_B + \mathcal{L}_M + \mathcal{L}_{\text{int}},   
\end{equation}
which take the explicit form
\begin{equation}
  \begin{split}
  \mathcal{L}_B =\,&  \bar{\Psi} ( \, i \gamma _\mu \partial^\mu - M ) \Psi,  \\
  \mathcal{L}_M =\,&  {\textstyle \frac{1}{2}} \sum_{\iota= \sigma, \delta} 
			\Big( \partial_\mu \Phi_\iota \partial^\mu \Phi_\iota - m_\iota^2 \Phi_\iota^2 \Big)   \\
  		 &	- {\textstyle \frac{1}{2}} \sum_{\kappa = \omega, \rho, \gamma }
			\Big( \textstyle{ \frac{1}{2}} F_{(\kappa) \mu \nu}\, F_{(\kappa)}^{\mu \nu}
				- m_\kappa^2 A_{(\kappa)\mu} A_{(\kappa)}^{\mu} \Big),      \\
  \mathcal{L}_{\text{int}} =\,&	- g_\sigma\bar{\Psi}  \Phi_\sigma \Psi 
                - g_\delta \bar{\Psi}  \boldsymbol{\tau} \boldsymbol{\Phi}_\delta \Psi \\
		& - g_\omega \bar{\Psi}  \gamma_\mu A_{(\omega)}^{ \mu } \Psi
		- g_\rho \bar{\Psi}  \boldsymbol{\tau } \gamma_\mu  \boldsymbol{A}_{(\rho)}^{\mu } \Psi \\
                & - e \bar{\Psi}\gamma_\mu {\textstyle \frac{1}{2}}(1+ \tau_3 ) A_{(\gamma)}^{\mu} \Psi ,
  \end{split}
\end{equation}
with the field strength tensor
$F_{(\kappa)\mu \nu} = \partial_{\mu} A_{(\kappa)\nu} - \partial_\nu A_{(\kappa)\mu}$
for the vector mesons.
In the above Lagrangian density the nucleon field consisting of Dirac--spinors in 
isospin space is denoted by $\Psi$ and the nucleon rest mass by $M = 938.9$ MeV.
The scalar meson fields are $ \Phi_\sigma$ and $ \boldsymbol{\Phi}_\delta $,
the vector meson fields $ A_{(\omega)} $ and $ \boldsymbol{A}_{(\rho)} $.
Bold symbols denote vectors in the isospin space acting between the two species of nucleons.
The mesons have rest masses $m_\kappa$ for each meson $\kappa$ 
and couple to the nucleons with the strength of the coupling constants $g_\kappa$.
The electromagnetic field $A_{(\gamma)}$ couples to the nucleons by the
electron charge $ e^2 = 4 \pi \alpha$ where $\alpha$
is the fine structure constant.
Notations are taken from \cite{Bjorken_Drell:1964}:
$x= x^\mu$ and $ x_\mu$ denote the contravariant and covariant 
vectors in space-time, $\gamma^\mu$ the Dirac $\gamma$ matrices and
$\boldsymbol{\tau}$ consists of the isospin Pauli matrices $\tau_k$.

Allowing for a density dependence of the baryon--meson vertices improves 
the capability of the model significantly, 
since it enables the latter to assimilate 
the self--energies of a Dirac--Brueckner--Hartree--Fock calculation.
This means that the coupling constants $g_\kappa$ 
depend on a density $\rho(\bar{\Psi}, \Psi)$ 
obtained from the nucleon field $\Psi$.
In the literature we find a dependence on the scalar density 
or on the vector density \cite{Fuchs:1995}.
It turned out that the dependence on the zero component of the vector density, 
the baryon density $\rho = \Tr( \bar{\Psi} \gamma_0 \Psi ) $, 
is the most suitable one since it describes finite nuclei better 
and has a natural connection to the vertices in the DBHF calculations \cite{Hofmann:2001}.
The functions for the various couplings
are specified later together with the associated parameter set.

Applying the variational principle to the Lagrangian
we obtain a Dirac equation for the nucleons and 
Klein--Gordon and Proca equations for the meson fields \cite{Fritz:1994}.

Due to density dependent vertices the variation principle changes to
\begin{equation}
  \frac{\delta \mathcal{L}}{\delta \bar{\Psi}} 
	= \frac{\partial \mathcal{L}}{\partial \bar{\Psi}}
	+ \frac{\partial \mathcal{L}}{\partial \rho} \frac{\delta \rho}{\delta \bar{\Psi}},
\end{equation}
where the second expression creates the so called rearrangement 
contribution $\Sigma_R$ to the self--energies of the nucleon field.
These rearrangement contributions contribute only to the zero component
of the vector self--energy.
Including these additional contributions we denote the Dirac equation
for the nucleonic single--particle wave function $\psi_\alpha$ 
in Hartree approximation 
\begin{equation}\label{eq:Dirac}
\big(\boldsymbol{\alpha} \boldsymbol{p} + (\Sigma_0 + \Sigma_R)
     + \beta (M + \Sigma_S) \big)\, \psi_\alpha
  = \epsilon_\alpha \, \psi_\alpha, 
\end{equation}
where the self--energy contributions read
\begin{align}\label{eq:self_en}
  \Sigma_S &= g_\sigma \Phi_\sigma + g_\delta \Phi_\delta \tau_3, \notag \\
  \Sigma_0 &= g_\omega A^{(\omega)}_0 + g_\rho A_0^{(\rho)} \tau_3 
	      + e\, \frac{1}{2} (1-\tau_3)A_0^{(\gamma)},
\end{align}
and the rearrangement self--energy contribution $\Sigma_R$ is obtained by
\begin{equation}\label{eq:self_en_rearr}
  \Sigma_R = \Big(
	  \frac{\partial g_\sigma}{\partial \rho} \Phi_\sigma \rho^s
	 + \frac{\partial g_\delta}{\partial \rho} \boldsymbol{\Phi}_\delta \rho_3^s
	 + \frac{\partial g_\omega}{\partial \rho} \gamma_\mu A^{(\omega)}_0 \rho
	 + \frac{\partial g_\rho}{\partial \rho} \boldsymbol{A}^{(\rho)}_0 \rho_3
            \Big).
\end{equation}

The various densities are obtained from the nucleon single--particle wave functions 
in the ''no--sea'' approximation as
\begin{eqnarray}\label{eq:density}
  \rho^s(\boldsymbol{x})	
	&=&	\sum_{\alpha} \eta_\alpha  \, \bar{\psi}_\alpha(\boldsymbol{x}) \psi_\alpha(\boldsymbol{x}) \notag \\
  \rho^s_3(\boldsymbol{x}) 	
  	&=&	\sum_{\alpha} \eta_\alpha  \, \bar{\psi}_\alpha(\boldsymbol{x}) 
			\tau_3 \psi_\alpha(\boldsymbol{x}) \notag \\
  \rho(\boldsymbol{x})		
  	&=&	\sum_{\alpha} \eta_\alpha  \, \bar{\psi}_\alpha(\boldsymbol{x}) \gamma_0 \psi_\alpha(\boldsymbol{x}) 
		\notag \\
  \rho_3(\boldsymbol{x})	
  	&=&	\sum_{\alpha} \eta_\alpha  \, \bar{\psi}_\alpha(\boldsymbol{x}) 
					      \gamma_0 \tau_3 \psi_\alpha(\boldsymbol{x})  \notag \\
  \rho^{(em)}(\boldsymbol{x})	
  	&=&	\sum_{\alpha} \eta_\alpha  \, \bar{\psi}_\alpha(\boldsymbol{x}) 
						{\textstyle \frac{1}{2}}(1-\tau_3) \psi_\alpha(\boldsymbol{x})
					    \ \  [- \rho_e(\boldsymbol{x})].
\end{eqnarray}
where $\rho^s$ is the scalar density, $\rho$ the baryon density, 
$\rho^s_3$ the scalar isovector density, $\rho_3$ the vector isovector density, 
and $\rho^{(em)}$ the charge density. The occupation factors $\eta_\alpha$ have to be determined
from the desired scheme of occupation.

Neglecting retardation effects 
the Klein-Gordon equations reduce to inhomogeneous Helmholtz equations 
with source terms \cite{Fritz:1994}
\begin{align}\label{eq:meson_fields}
  ( - \Delta + m_\sigma^2)\, \Phi_\sigma	
  	&= - g_\sigma  \, \rho^s  \notag \\
  ( - \Delta + m_\delta^2)\, \Phi_\delta
        &= - g_\delta \, \rho^s_3 \notag \\
  ( - \Delta + m_\omega^2)\, A_0^{(\omega)}
  	&= \quad	g_\omega \, \rho \notag \\      
  ( - \Delta + m_\rho^2)\, A_0^{(\rho)}
  	&= \quad	g_\rho  \, \rho_3  \notag \\
    - \Delta \, A_0^{(\gamma)}
    	&= \quad	e \, \rho^{(em)},
\end{align}
from which the self--energy contributions (\ref{eq:self_en}) are obtained.
The Dirac equation for the nucleons (\ref{eq:Dirac}), the evaluation of the
resulting densities (\ref{eq:density}), these meson field equations 
(\ref{eq:meson_fields}) and the calculation of the resulting self--energy
contributions(\ref{eq:self_en}) form a set of equations, which have to be solved
in a self--consistent way.

\subsection{Nuclear Matter}

In nuclear matter the electromagnetic field is neglected and translational
invariance is assumed. 
The Dirac spinors $u(k,s,i)$ with momentum $k$, spin $s$ and isospin $i$
are solutions of the following Dirac equation:
\begin{equation}
  [\boldsymbol{\gamma} \cdot \boldsymbol{k}^\ast + m_{i}^\ast] u(k,s,i) = \gamma_0 E_i^\ast u(k,s,i),
\label{eq:NM_Dirac}
\end{equation}
where the starred quantities are defined by 
\begin{eqnarray}
  m_i^\ast 
  	&=& M + \Sigma_{S,i}(k),     \notag	\\
  E_i^\ast &=& E(k) - \Sigma_{0,i} (k), 
\end{eqnarray}
and the on-shell condition reads 
\begin{equation}
  {E^\ast_i}^2 = \boldsymbol{k}^2 + {m_i^\ast}^2.
\end{equation}
The positive energy solutions of the Dirac equation (\ref{eq:NM_Dirac}) are
obtained in straight analogy to the results for the free case by
\begin{equation}
  u(k,s,i) 
  	= \left( \frac{E^\ast_i + m^\ast_i}{2E^\ast_i} \right)^{1/2} 
  	  \begin{pmatrix}
	   1 \\
	   \frac{\boldsymbol{\sigma} \cdot \boldsymbol{k_i^\ast}}{E^\ast_i + m^\ast_i} 
	  \end{pmatrix}
	  \chi_{1/2}(s) \chi_{1/2}(i),
\end{equation}
where the spinors $\chi_{1/2}(s)$ and $ \chi_{1/2}(i)$ account for spin and isospin projection.

From these spinors the vector and scalar densities for protons and neutrons
can easily be calculated in infinite matter as
\begin{eqnarray}\label{eq:NM_dens}
  \rho^s_i &=& \frac{1}{\pi^2} \int \eta_i(k)\, k^2 \, dk \,  \frac{m_i^\ast}{E_i(k)^\ast},  \notag \\
  \rho_i &=& \frac{1}{\pi^2} \int \eta_i(k)\, k^2 \, dk.
\end{eqnarray}
The energy density $\mathcal{E}$ and the pressure $P$ are obtained from the 
energy--momentum tensor
\begin{eqnarray}
\mathcal{E} = \langle T^{00} \rangle
	&=& \frac{1}{\pi^2} \sum_{i=p,n} \int \eta_i(k)\, k^2 \, dk \,
		\sqrt{ k^2 + m_i^{\ast^2} }		\notag		\\
	&& + \frac{1}{2} \sum_{i=p,n} \left( \Sigma_{S,i} \, \rho_i^s + \Sigma_{0,i} \, \rho_i \right)	\\
P = \frac{1}{3} \sum_{j = 1}^3 \langle T^{jj} \rangle
	&=&  \frac{1}{3\pi^2} \sum_{i=p,n} \int \eta_i(k)\, dk \,
		\frac{k^4}{\sqrt{ k^2 + m_i^{\ast^2} }}	\nonumber \\
	&& + \Sigma_R\, \rho
	   + \frac{1}{2} \sum_{i=p,n} \left( -\Sigma_{S,i} \, \rho_i^s + 
	   \Sigma_{0,i} \, \rho_i \right).\label{pressure}
\end{eqnarray}
Rearrangement does not contribute to the energy density but it modifies the pressure, which
coincides with the corresponding thermodynamical relation.

The entropy density $\mathcal{S}$ is obtained as
\begin{equation}
  \mathcal{S} = - \frac{1}{\pi^2} \sum_{i=p,n} \int \eta_i(k)\, k^2 \, dk 
		  \left[ f_i(k) \, \ln(f_i(k)) + (1-f_i(k) ) \, \ln(1-f_i(k)) \right],
\end{equation}
where $f_i(k)$ are the thermal occupation probabilities.
Note that $\eta_i$ and $f_i$ may differ, e.g. if 
pairing is introduced in the finite temperature BCS formalism.
The energy density $\mathcal{E}$ and the entropy density $\mathcal{S}$ 
yield the free energy density
\begin{equation}
  \mathcal{F} = \mathcal{E} - T \mathcal{S}
\end{equation}
and the thermodynamic potential
\begin{equation}
  \Omega = \mathcal{E} - T\mathcal{S} - \mu_p \rho_p - \mu_n \rho_n.
\end{equation}
Minimizing the thermodynamic potential $\Omega$ for fixed meson fields
we obtain the chemical potentials 
\begin{equation}
  \mu_i = \varepsilon_{F,i} + \Sigma_{0,i} + \Sigma_R
\end{equation}
from the Fermi energy $\varepsilon_{F,i} = E_i^\ast(k_{F,i})$
and the thermal occupation factors in the 
''no-sea'' approximation
\begin{equation}
  f_i (k) = \frac{1}{1+ \exp[ (E_i^\ast - \varepsilon_{F,i}) / T ]}, \ \ \ i = p,n.
\end{equation}
The Fermi momenta $k_{F,i}$ for protons and neutrons have to be determined
from the given proton and neutron densities (see eq. \ref{eq:NM_dens}).
A self--consistent calculation including densities, Fermi momenta,
effective masses and the self--energy contributions for protons and neutrons
yields all described quantities.

\subsection{Finite Systems}

For finite nuclei and the description of nuclear matter in a Wigner--Seitz cell 
(WS) the Dirac equation (\ref{eq:Dirac}) and the meson field equations (\ref{eq:meson_fields})
are solved in spatial representation.
The numerical procedure to solve the Dirac equation in the cubic box is the same as
in \cite{Goegelein:2007} with an extension of the code to allow for the
density-dependent coupling constants.
Pairing correlations are included like in \cite{Montani:2004, Goegelein:2007}.
To be able to perform consistent finite temperature calculations 
the finite temperature BCS formalism (FT-BCS) has been included in the code 
which allows the simultaneous evaluation of pairing and finite
temperature effects. 

According to Goodman \cite{Goodman:1981} the 
normal and anomalous occupation factors of the Hartree--Fock single particle states
are obtained in the FT-BCS approach as
\begin{eqnarray}\label{eq:therm_occs}
  \eta_\alpha &=& (1 - 2 f_\alpha) v_\alpha^2 + f_\alpha,   \notag	\\
  \zeta_\alpha &=& (1 - 2 f_\alpha) \, u_\alpha v_\alpha,
\end{eqnarray}
where $v_\alpha$ and $u_\alpha$ denote the usual quasi--particle occupation
factors and $f_\alpha$ the thermal occupation factors
\begin{equation}
  f_\alpha = \frac{1}{1+ \exp[ E_\alpha / T]}
\end{equation}
depending on the quasi--particle energy 
$E_\alpha = \sqrt{(\varepsilon_\alpha - \varepsilon_{F,i})^2 + \Delta_\alpha^2 }$,
where $\varepsilon_\alpha$ are the single--particle energies and $\Delta_\alpha$ the
state--depending pairing gap.

The normal occupation $\eta$ modifies the densities (\ref{eq:density}) and (\ref{eq:NM_dens}), 
while the anomalous occupation factors enter into the anomalous density
\begin{equation}\label{eq:anomalous_dens}
  \chi (\mathbf{r}) = {\textstyle \frac{1}{2}} 
  			\sum_{\alpha}
			 \, \zeta_\alpha
			 \left| \psi_\alpha(\mathbf{r}) \right|^2.
\end{equation}
The local gap function based on a density dependent pairing force $V$ of zero
range  
\begin{equation}\label{eq:local_gapfunc}
  \Delta (\mathbf{r}) = - V(\mathbf{r}) \, \chi(\mathbf{r}),
\end{equation}
modifies the state--dependent pairing gaps
\begin{equation}
 \Delta_\alpha = \int d^3\mathbf{r} \  
	\Delta (\mathbf{r})\,  \left| \psi_\alpha(\mathbf{r}) \right|^2.
\end{equation}
These pairing gaps $\Delta_\alpha$ are evaluated in a 
self--consistent procedure fixing the Fermi energies for protons and neutrons
$\varepsilon_{F,i}$
by the corresponding particle number conditions
\begin{equation}
  N = \sum_\alpha \eta_\alpha.
\end{equation}
Finally the the pairing energy $E_{pair}$ 
 is obtained from the state dependent pairing gaps
\begin{equation}
  E_{pair} = \frac{1}{2} \sum_{\alpha} \Delta_\alpha \zeta_\alpha.
\end{equation}

The center of mass correction which gives a significant contribution to the binding energy of 
light nuclei is treated in the usual harmonic oscillator approximation like in \cite{Hofmann:2001}
\begin{equation}
  E_{cm} = - \frac{3}{4} \hbar \omega 
\end{equation}
with $\hbar \omega = 41\, A^{-1/3}$ MeV.

The total ground state energy includes the center of mass correction,
the pairing energy and the energy of the relativistic mean field 
\begin{equation}
  E_0 = E_{RMF} + E_{pair} + E_{cm},
\end{equation}
where the Hartree ground state energy is obtained from the Dirac Hamiltonian 
corresponding to the Lagrangian density (\ref{Lag_dens}) similar to \cite{Fritz:1994}. 
Considering the rearrangement and the single--particle energies
the ground state energy is calculated by
\begin{equation}
\begin{split}
  E_{RMF} = & \sum_\alpha \eta_\alpha \varepsilon_\alpha 
		- \int d^3r \, \Sigma_R(\boldsymbol{r}) \, \rho(\boldsymbol{r}) \\
	& - \frac{1}{2} \sum_{i=p,n} \int d^3r 
		\left[	\Sigma_{S,i}(\boldsymbol{r})   \, \rho^s_i(\boldsymbol{r}) 
			+ \Sigma_{0,i}(\boldsymbol{r}) \, \rho_i (\boldsymbol{r}) \right].
\end{split}
\end{equation}

\section{Density Dependent Parameterization from DBHF Theory}
\label{sec:Param_DBHF}

In this section we will describe the parameterization of density dependent
coupling constants to be used in the DDRMF approach, which is based on the DBHF 
calculations for homogeneous asymmetric nuclear matter of van Dalen et
al.\cite{vanDalen:2007}, which employ the meson exchange interaction model OBEPA
of the Bonn potentials\cite{brock}. The aim is to obtain a parameterization,
which accurately reproduces the results of the DBHF calculations at high
densities but also provides a good description of bulk properties of finite
nuclei. This should lead to an equation of state covering a very broad range of
densities as e.g. in Refs. \cite{Shen:1998a, Shen:1998b}.

A first attempt to translate the results of DBHF calculations in terms of
density-dependent coupling constants would be to consider the scalar ($\Sigsi$)
and vector contributions ($\Sigoi$) for protons and neutrons calculated in the
DBHF approximation for asymmetric nuclear matter of a density $\rho$ and proton
- neutron asymmetry $\rho_3$ and equate these self-energy terms with the
corresponding mean-field expressions. This leads to the following expressions
for the effective coupling constants for the $\sigma$, $\omega$, $\delta$, 
and $\rho$ mesons\footnote{The coupling constants of effective relativistic
field theories $g_\kappa$ are generally presented as dimensionless quantities. 
These dimensionless quantities are obtained from the expressions in
eqs. (\ref{eq:ss})-(\ref{eq:ivr})
by dividing all values for masses and self-energies by $\hbar c$}:
\begin{eqnarray}
\left(\frac{g_{\sigma}(\rho ,\rho_3)}{m_{\sigma}}\right)^2 = - \frac{1}{2}
\frac{\Sigsp(k_{\mathrm Fp})  + \Sigsn(k_{\mathrm Fn})}{\rho^s}, \label{eq:ss}\\
\left(\frac{g_{\omega}(\rho ,\rho_3)}{m_{\omega}}\right)^2 = - \frac{1}{2}
\frac{\Sigop(k_{\mathrm Fp})  + \Sigon(k_{\mathrm Fn})}{\rho }, \label{eq:vw} \\
\left(\frac{g_{\delta}(\rho ,\rho_3)}{m_{\delta}}\right)^2 = - \frac{1}{2}
\frac{\Sigsp(k_{\mathrm Fp})  - \Sigsn(k_{\mathrm Fn})}{\rho^s_{3}}, \label{eq:isd}\\
\left(\frac{g_{\rho}(\rho ,\rho_3)}{m_{\rho}}\right)^2 = - \frac{1}{2}
\frac{\Sigop(k_{\mathrm Fp})  - \Sigon(k_{\mathrm Fn})}{\rho_3},  \label{eq:ivr}
\end{eqnarray}
with $\rho^s =\rho^s_{p}+ \rho^s_{n}$, $\rho =\rho _p+\rho _n$,
$\rho^s_{3}=\rho^s_{p}-\rho^s_{n}$, and $\rho _3=\rho _p- \rho _n$,
where
\begin{eqnarray}
\rho^s_{i}=\frac{2}{(2\pi)^3} \int_0^{k_{Fi}} d^3{k} \frac{m^*_i}{\sqrt{{m^*_i}^2+k^2}}
\end{eqnarray}
and
\begin{eqnarray}
\rho _i=\frac{2}{(2\pi)^3} \int_0^{k_{Fi}} d^3{k} = \frac{k_{Fi}^3}{3 \pi^2}
\end{eqnarray}
are, respectively, the scalar and vector density of particle $(i=n,p)$. In
contrast to widely used RMF theories we explicitly include the scalar isovector
meson $\delta$ since this provides a mechanism to account for the differences in
the scalar self-energies and the corresponding effective Dirac masses for 
protons and neutrons in asymmetric matter\cite{Schiller:2001}.  Following
Ref.~\cite{Dejong:1998} the inclusion of the $\delta$ meson in the  DDRMF theory
has important consequences for the dynamics of neutron-rich nuclei. In addition,
this meson is important for astrophysical applications since the dense
asymmetric matter gets softer, which means that the pressure rises more slowly
for larger densities.

The expressions of eqs. (\ref{eq:ss})-(\ref{eq:ivr}) identify coupling
constants, which depend on the density $\rho$ and the proton - neutron asymmetry
($\rho_3$) Since, however, the dependence of the
coupling functions turns out to be weak (see  e.g. 
\cite{Schiller:2001,vanDalen:2004b,vanDalen:2007}), we have ignored this
dependence on $\rho_3$ to keep the DDRMF functional as simple as possible.

The fact that a kind of renormalization is required when DBHF results are mapped
on Relativistic Mean Field (RMF) theory has already been pointed out in Ref.
\cite{Hofmann:2001,vanDalen:2007}.  The reason is that two essential differences
exist between DBHF and RMF concerning the structure of the self energy. First,
the DBHF self energy terms explicitly depend on the momentum of the particle, a
feature which is absent in RMF. This reflects the non-locality of the DBHF
self-energy terms, which originates from the Fock exchange terms but also from
non-localities in the underlying NN interaction. Since this explicit momentum
dependence of the DBHF self energy components is generally
weak~\cite{Gross:1999,vanDalen:2004b}, we will ignore this non-locality effects
and always consider the DBHF self-energy terms calculated  at the corresponding
Fermi momenta $k_{Fi}$ as already expressed in eqs. 
(\ref{eq:ss})-(\ref{eq:ivr}).
 
Secondly, the appearance of a spatial contribution of the vector self energy
$\Sigma_V$ in the DBHF theory, which is not present in the RMF model.  The
$\Sigma_V$ component originates from Fock exchange contributions which are not
present in the RMF theory. For an accurate reproduction of the DBHF energy
functional the spatial $\Sigma_V$ component has to be included in a proper way.
The effects of the the $\Sigma_V$ component in the Dirac equation for
homogeneous nuclear matter can be absorbed into a renormalization of the scalar
and time like vector component of the self-energy. This leads to an
effective Dirac mass, which has to be identified with the RMF effective mass, i.e.
\begin{eqnarray}
\tilde{m}^{*}_i=\frac{M + \Sigsi(k_{Fi})}{1+\Sigvi(k_{Fi})}=M+\Sigma^{RMF}_{{\mathrm s},i}.
\label{eq:meffDBHF}
\end{eqnarray}
This leads to the renormalized scalar self energy component
\begin{eqnarray}
\Sigma^{RMF}_{{\mathrm s},i}=\frac{\Sigsi(k_{Fi})-M \Sigvi(k_{Fi})}{1+\Sigvi(k_{Fi})}.
\end{eqnarray}
In a corresponding way the following expression for the renormalized vector self
energy component is obtained
\begin{eqnarray}
\Sigma^{RMF}_{{\mathrm 0},i}=\Sigoi(k_{Fi})-\frac{\Sigvi(k_{Fi}) [3 E_{Fi} \rho_i 
		+  \tilde{m}^*_i \rho_{i}^s]}{4 \rho_i}.
\end{eqnarray}
These renormalized self-energy components are now inserted into eqs.~(\ref{eq:ss})-(\ref{eq:ivr})
to obtain the renormalized density dependent coupling functions.

The density dependent isoscalar couplings
are obtained from eqs.~(\ref{eq:ss})-(\ref{eq:ivr}) using the symmetric nuclear data ($\beta=0.0$)
up to a density of $0.5$ fm$^{-3}$,
whereas the isovector ones are obtained using the neutron matter data 
of the DBHF calculations in Ref. \cite{vanDalen:2007}.

In order to make this parameterization easily accessible, we have fixed the
masses of the mesons to the corresponding masses in the underlying free NN
interaction (see table \ref{table:DDRH_parameters_vD}) and parameterized 
the density dependence of the coupling constants by
\begin{equation}
  g_\kappa (\rho _B)
    = a_\kappa + \big[ b_\kappa + d_\kappa  x^3  \big]
       \exp(-c_\kappa x ),
\label{eq:EvD_coupl_func}
\end{equation}
where $x = \rho _B/\rho_0$, and $\rho_0$ = 0.16 fm$^{-3}$ is the 
saturation density of symmetric nuclear matter.
The values obtained for the fit of the
coupling functions are summarized in table \ref{table:DDRH_parameters_vD}.

\begin{table}
\begin{center}
\begin{tabular}{|c|cc|ccccc|}
\hline
\ \ \ $\kappa$ &	\ $J^P$ \ & \ I \ & \ $m$ [MeV]\ \ \ &\ \ \ $a_\kappa$\ \ \ &\ \ \ $b_\kappa$ \ \ \ 
& \ \ \ $c_\kappa$ \ \ \ & \ \ \ $d_\kappa$ \ \ \  	\\
\hline
 $\sigma$ & $0^+$  & 0	& 550 & $7.7868$     & $2.58637$    & $2.32431$   &  $3.11504$   \\
\hline
 $\omega$ & $1^-$	& 0	& 782.6 & $9.73684$    & $2.26377$    & $7.05897$   &   -   \\
\hline
 $\delta$ & $0^+$	& 1	& 983 & $0.503759$   & $2.68849$    & $6.7193$    & $0.403927$ \\
\hline
 $\rho$   & $1^-$	& 1	& 769   & $4.56919$    & $5.45085$    & $1.20926$   &  -    \\
\hline
\end{tabular}
\end{center}
\caption{\label{table:DDRH_parameters_vD}Parameter set from DBHF by van Dalen et
	 al. \cite{vanDalen:2007} for the density dependent relativistic mean
	 field approach.}
\end{table}

Employing these functions in a DDRMF calculation of finite nuclei, we obtained
binding energies and radii, which are too small as compared to the empirical
data. This is in line with the observation that also the underlying DBHF 
calculations for symmetric nuclear matter yield a saturation density, which is
too large as compared to the empirical result (see table
\ref{tab:NM_prop_vD_param}). In order to cure these problems, we have considered
a slight reduction of the coupling constant for the $\omega$ meson at densities
around the saturation density in a form:  
\begin{equation}
  g_{\omega, cor}(\rho _B) =
    g_\omega(\rho) - a_{cor}
       \exp\Big( - \Big[ \frac{\rho - \rho_0}{b_{cor}} \Big]^2 \Big).
\label{eq:EvD_coupl_corr}
\end{equation}
Adjusting the parameters of this correction to $a_{cor}$ = 0.014 and $b_{cor}$ =
0.035 fm$^{-3}$, we obtain an improvement for the saturation density of nuclear matter
(see table \ref{tab:NM_prop_vD_param}) a rather good description for energies
and radii of finite nuclei (see table \ref{tab:FN_vD_param}).
The lighter nuclei are a little
too much bound but the binding energies for
heavy nuclei are well reproduced.
The charge radii $r_c$ show a good agreement
for the lighter nuclei whereas for the heavy nuclei
the radii are a little too small.

\begin{table}
\begin{center}
\begin{tabular}{|lc|cc|}
\hline
	     &   	& DBHF 		& DDRMF    	\\
\hline
$\rho_{sat}$ & [MeV] 	& $0.181$	& $0.166$	\\
$E/A $       & [MeV]    & $-16.15$	& $-16.23$	\\
$K$	     & [MeV]    & $230$		& $335$		\\
$a_s$	     & [MeV]    & $34.36$	& $29.92$	\\
\hline
\end{tabular}
\end{center}
\caption{\label{tab:NM_prop_vD_param}
	 Nuclear matter properties obtained in the different models.
	 The DDRMF model includes the correction from renormalization.}
\end{table}

\begin{table}[h]
\begin{center}
\begin{tabular}{|lc|ccccc|}
\hline
	&	 & \ $^{16}O$ \ & \ $^{40}Ca$ \ & \ $^{48}Ca$ \ & \ $^{90}Zr$ \ & \ $^{208}Pb$ \   \\
\hline
$E/A$   & [MeV]	 & $-8.35$	& -8.73		& -8.73		& -8.74		& -7.87	   \\
$E/A$ exp. & [MeV] & -7.98	& -8.55		& -8.67		& -8.71		& -7.87	   \\
\hline
$r_c$   & [fm] 	 & 2.78		& 3.44		& 3.45		& 4.17		& 5.31     \\
$r_c$ exp. & [fm] & 2.74	& 3.48		& 3.47		& 4.27		& 5.50     \\
\hline
\end{tabular}
\end{center}
\caption{\label{tab:FN_vD_param}
          Results for closed shell nuclei applying the
	  DDRMF parameterization from the DBHF results.
	  Experimental values are taken from \cite{Hofmann:2001}.}
\end{table}

In table \ref{tab:NM_prop_vD_param} the nuclear matter properties of the DBHF
calculations are compared to those of the DDRMF using the adjustment of
$g_\omega$ according to eq.(\ref{eq:EvD_coupl_corr}). Due to this adjustment the
nuclear matter properties are slightly changed around the saturation point. The
saturation density is shifted to a lower value so it is even closer to the
experimental value. The energy per nucleon $E/A$ of symmetric nuclear matter is
fairly well reproduced but the compression modulus $K$ is rising due to the fit
procedure. In this context it is worth to mention that RMF fits to finite nuclei
require relatively high compression modulus K of about 300 MeV \cite{Ring:1996}.  EoSs
with a stiff high density behavior stand, however, in contrast to the
information extracted from heavy ion reactions \cite{Sturm:2001}. Note, however,
that the correction for the $\omega$ coupling function is restricted to
densities around $\rho_0$ and should not affect the high density behavior of the
EoS.

The symmetry energy coefficient $a_s$ is decreased by about 4 MeV compared to
the DBHF results. This reduction of the symmetry energy is connected to the 
decrease of the saturation density as the two values are read of at 
essentially different densities. 
These changes should not affect the
astrophysical properties of the model in the high density region, what is
important to be able to reproduce the heavy neutron stars.
Moreover, the value of $a_s\simeq 30$ MeV at $\rho_{sat} = 0.166~{\rm fm}^{-3}$ 
agrees well with the presently favored experimental value of 
$a_s\simeq 31\pm 1$ MeV \cite{Khoa:2005ze}.

\section{Results and Discussion}


In this section we are going to discuss results of
density dependent relativistic mean field calculations (DDRMF) 
employing the parameterization of effective coupling constants as
described in subsection (\ref{sec:Param_DBHF}).
Pairing correlations are included in terms of the BCS approximation assuming a 
density dependent zero-range pairing force, which has 
already been used in earlier investigations \cite{Montani:2004, Goegelein:2007}.

The calculations are performed in a cubic Wigner
Seitz (WS) cell. 
The origin of the coordinate system is put in the center of the box and we
assume the density profiles to be symmetric under reflection on the $x=0$, $y=0$
and $z=0$ planes. Therefore we can restrict the calculation to a cubic box of
length $R$ in each direction,
so that the ''size'' of this box $R$ in each direction is half the length of 
the WS cell.
This box size $R$ has been adjusted to minimize the 
total energy per nucleon for the density under consideration.

All calculations which we discuss in this section for charge neutral matter 
containing protons, electrons and neutrons in $\beta$--equilibrium. This implies
that the chemical potentials of these species obey
\begin{equation}
     \mu_n = \mu_p + \mu_e.
\end{equation}

\begin{figure}
\begin{center}
  \mbox{ \includegraphics[width = 11cm]{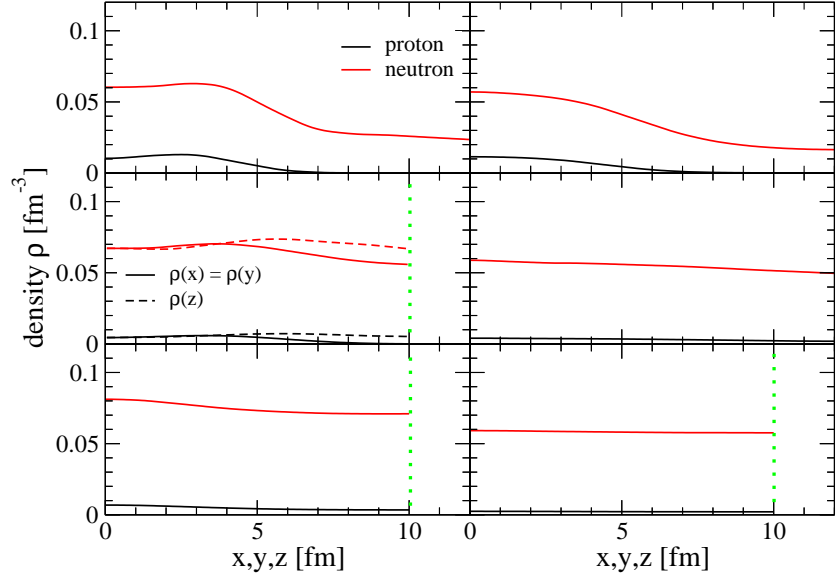} }
\end{center}
\caption{\label{fig:shapesDDRMF} (Color online)
Density distributions resulting from density dependent relativistic mean field
(DDRMF) calculations for protons (black color) and neutrons (red color) as a function 
of Cartesian coordinates $x,y,z$. The panels in the left column refer to
zero temperature calculations at the 
densities 0.020 fm$^{-3}$ (top), 0.057 fm$^{-3}$, and 0.067 fm$^{-3}$
(bottom), while those in the right column are obtained for the temperature
$T=5$ MeV at the densities
0.018 fm$^{-3}$ (top), 0.044 fm$^{-3}$, and 0.057 fm$^{-3}$.}
\end{figure}

We start our discussion by exploring the existence of geometrical structures in
the density distribution of protons and neutrons, which are typical for the 
``pasta-phase'' in the crust of neutron stars. For that purpose we display such
density distributions obtained from DDRMF calculations at zero temperature in
the panels on the left hand side of Fig.~\ref{fig:shapesDDRMF}.

The top panel in the left column represents a
nuclear structure at a baryonic density of 0.020 fm$^{-3}$ at zero temperature. 
In this case the density profiles are identical in all three Cartesian directions.
In the center a quasi--nuclear of the WS cell droplet structure is formed 
with a neutron sea between these quasi--nuclei.

\begin{figure}
\begin{center}
\protect{\vspace{10mm}}
\mbox{
  \includegraphics[width=6cm]{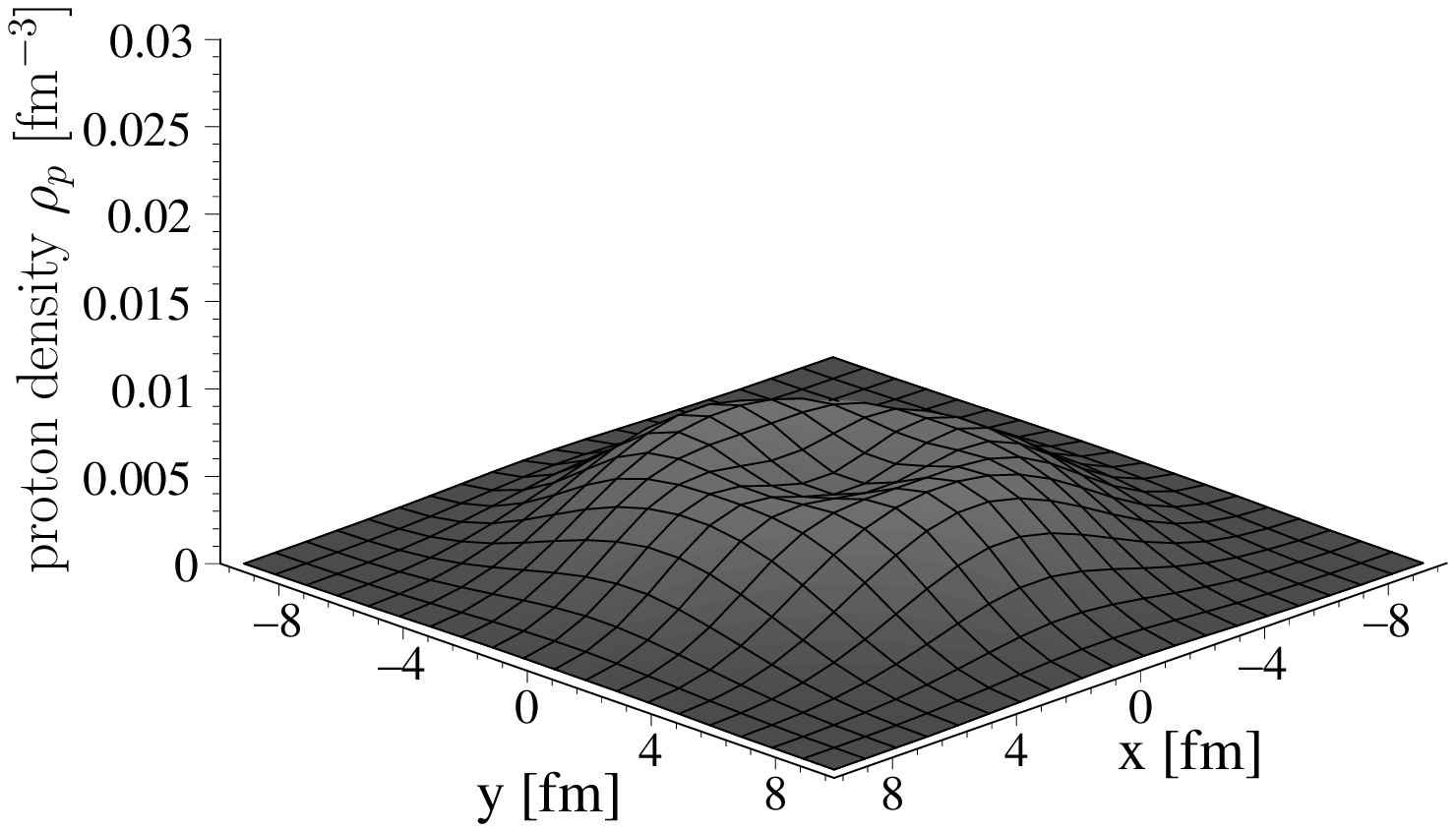}
\quad
  \includegraphics[width=6cm]{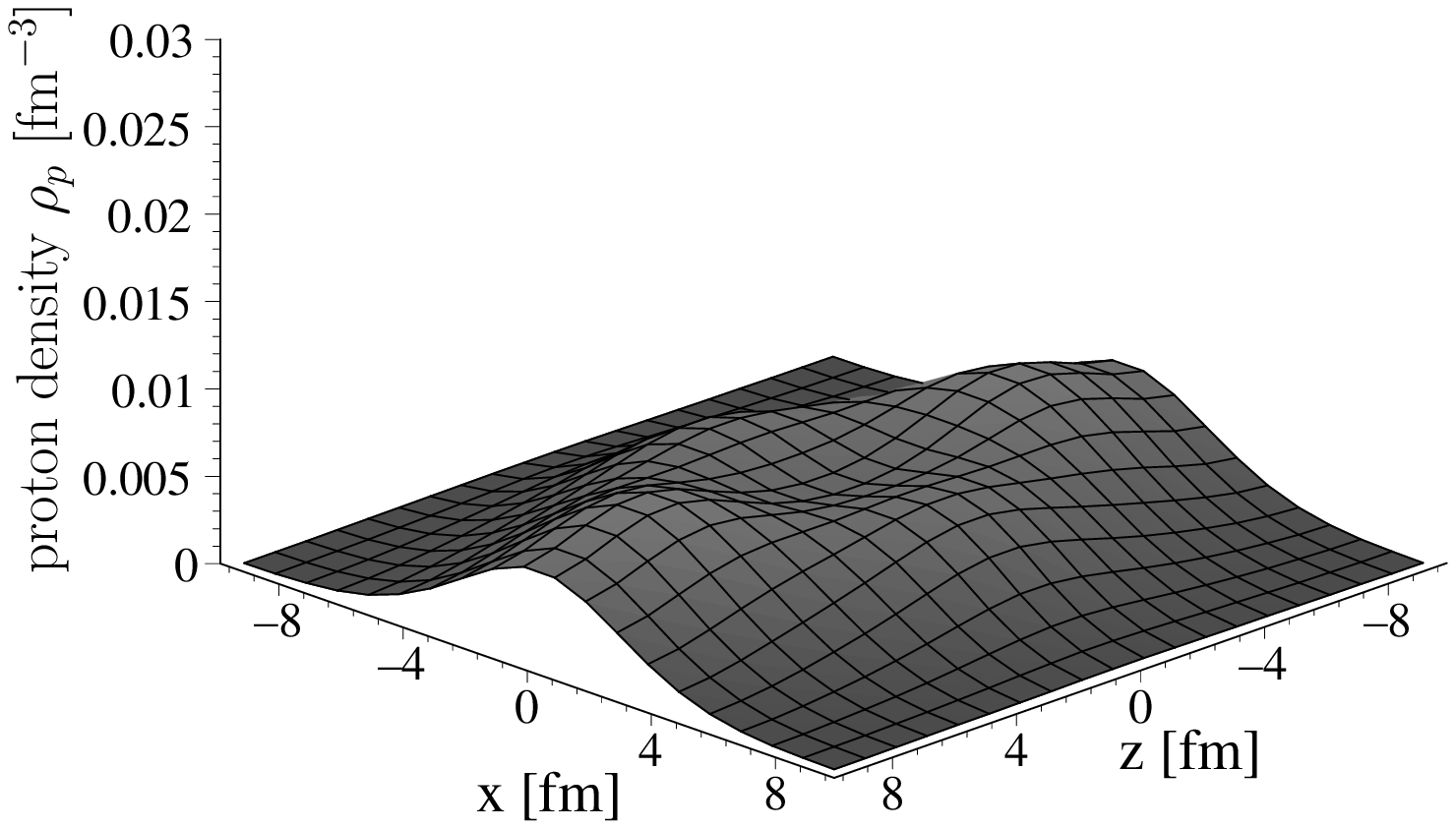}
}
\end{center}
\caption{\label{fig:DDRMFrod} 
	 Proton density distribution in the $z=0$ and the 
	 $y=0$ plane for the density dependent relativistic mean field (DDRMF)
	 calculation at an average density of $\rho = 0.057$ fm$^-3$.}
\end{figure}

Increasing the baryonic density to $\rho$ = 0.057 fm$^{-3}$, we obtain 
the second panel in the left column of Fig.~\ref{fig:shapesDDRMF}.  In
this case the proton density shows almost the same value along one axis, 
which is chosen to be the $z$--direction (dashed curves), while it drops to
zero, if the values for $x$ and $y$ tend towards $R$.
Here we find a so--called rod or Spaghetti-shape structure for the proton 
density distribution. Note, however, that the density distribution is not as
simple as these names rod-structure or Spaghetti-structure suggest. This can be
seen from the complete density-distribution in the $y=0$ (or $x=0$, which is
identical for this geometry), as it is displayed in Fig.~\ref{fig:DDRMFrod}.
There is a dip in the center of the quasi-nuclear structure and the density
distributions exhibit a structure, which is close to those of nuclei with 
strong
prolate deformations, which are aligned and touch each other along the $z$-axis.

\begin{figure}
\begin{center}
\protect{\vspace{10mm}}
\mbox{
  \includegraphics[width=6cm]{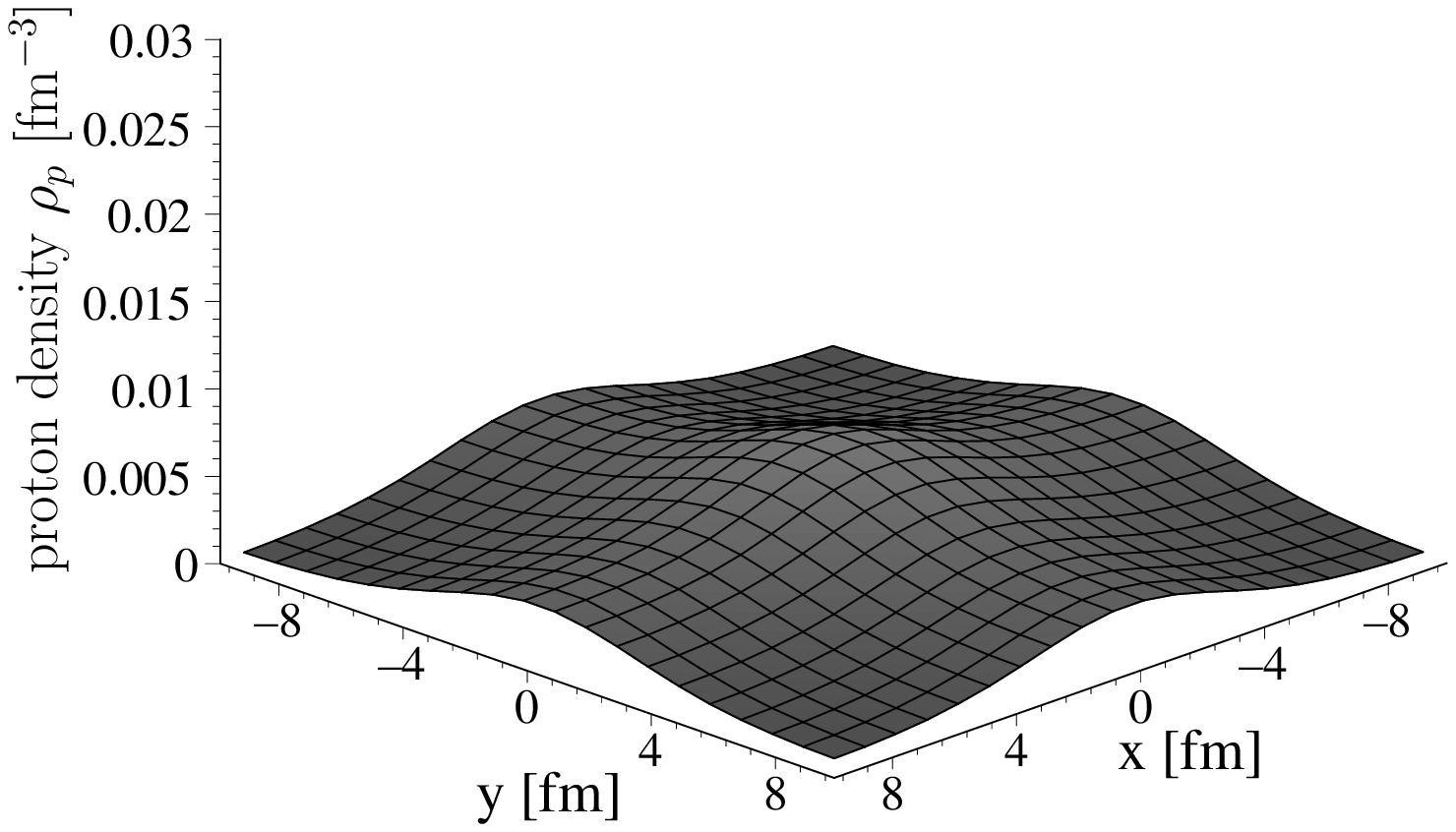}
\quad
  \includegraphics[width=6cm]{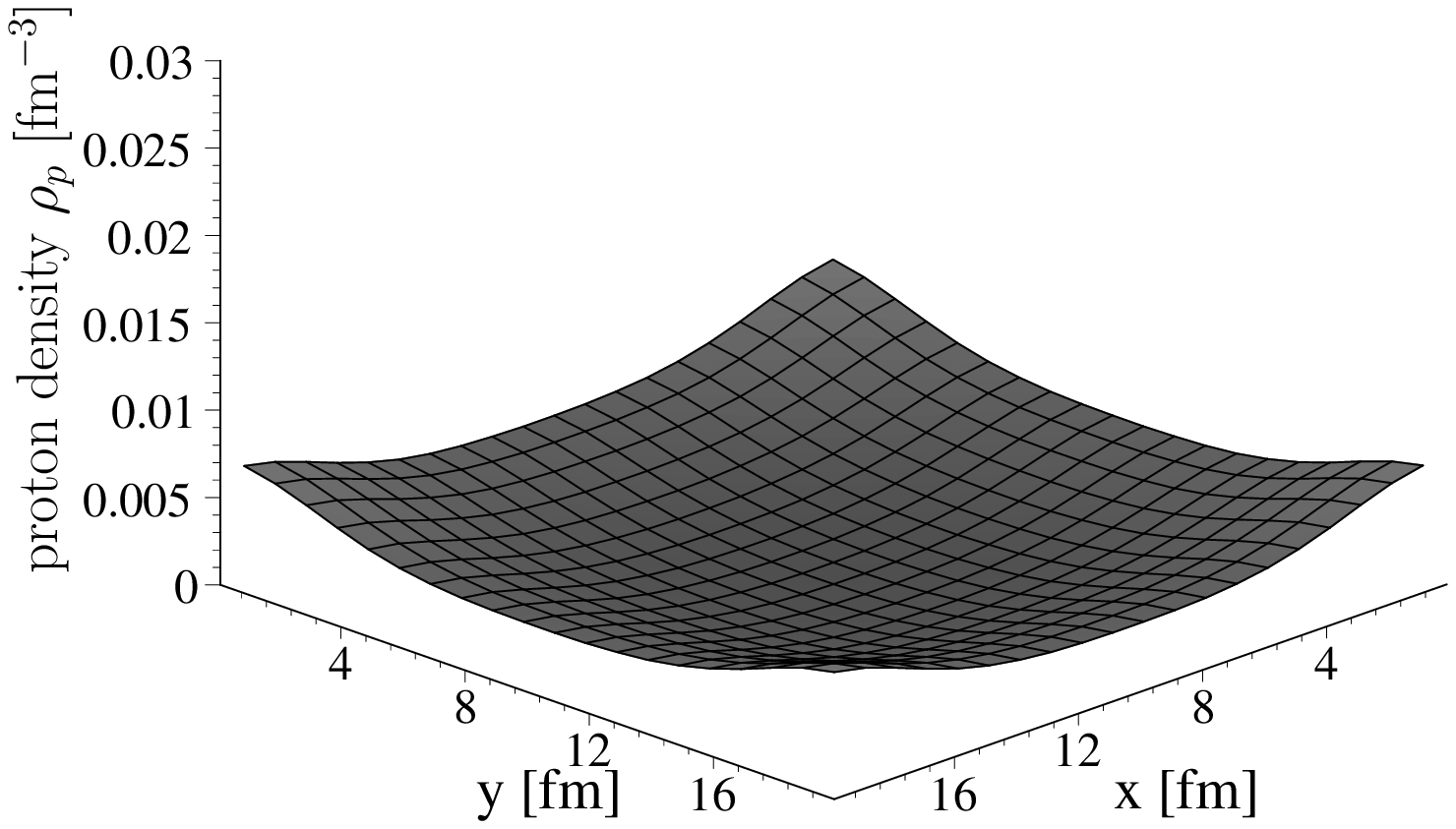}
}
\end{center}
\caption{\label{fig:DDRMFgrid} 
	 Proton density distribution in the $z=0$ plane 
	 for the density dependent relativistic mean field (DDRMF)
	 calculation at an average density of $\rho = 0.067$ fm$^-3$.
	 The surface plot on the left shows the density distribution 
	 with the origin in the center, while in one on the right hand side
	 the origin is shifted to the corner.
        }
\end{figure}

At larger densities the density distributions show again  the same distribution
along the axis in all Cartesian directions.  It is still slightly enhanced in
the center  and drops only by a small amount from the center to the boundary of
the cell. An example of such a structure  at the baryon density of 0.067
fm$^{-3}$ is displayed in  Fig.~\ref{fig:shapesDDRMF} in the bottom panel of the
left column. However, the proton density distribution in the $z=0$ plane 
displayed on the left in Fig.~\ref{fig:DDRMFgrid} provides a more detailed
picture of the geometric structures at this density: The proton density drops
along the diagonals, which implies that the regions of high densities in the center
of the WS cells are connected by arms along the Cartesian axes.  From this point
of view this structure may be called a grid structure. If, however,  the
coordinates are shifted in such a way, that the centers of the quasi-nuclei are
in the corners of the box as it is displayed in Fig.~\ref{fig:DDRMFgrid} on the
right, one finds that this structure can also be interpreted as  a bubble
structure. Summarizing we observe a grid structure with bubbles in between,
which provides a last step of a smooth transition to homogeneous matter.

\begin{figure}
\begin{center}
  \mbox{ \includegraphics[width=13cm]{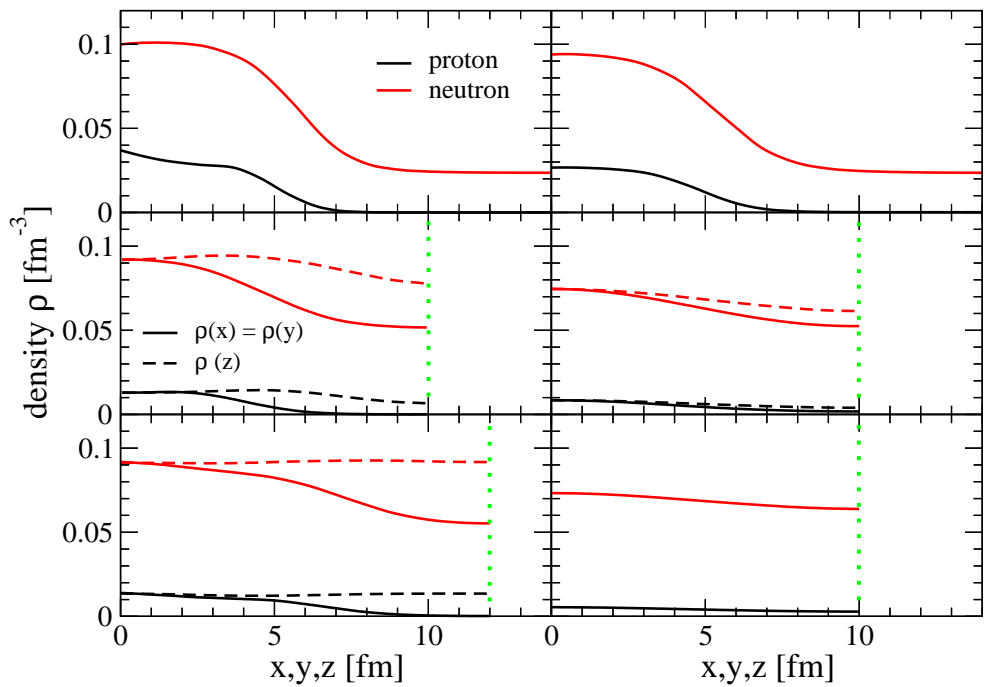} }
\end{center}
\caption{\label{fig:shapes_S5} (Color online)
Comparison of density distributions 
at finite temperature with those 
at zero temperature resulting from Skyrme HF
calculations for protons (black color) and neutrons (red color) as a function 
of Cartesian coordinates $x,y,z$. 
The panels in the left column refer to
$T=0$ MeV calculations at the average
densities 0.0271 fm$^{-3}$ (top), 0.0565 fm$^{-3}$, and 0.0625 fm$^{-3}$
(bottom), while those in the right column are obtained for 
$T=5$ MeV at the baryon densities
0.0273 fm$^{-3}$ (top), 0.0545 fm$^{-3}$, and 0.0648 fm$^{-3}$.}
\end{figure}

These kind of geometrical structures and others have also been obtained in 
a recent investigation employing
the Skyrme Hartree--Fock approach \cite{Goegelein:2007}. Some typical results
for the density profiles at  zero temperature are presented in the left column 
of Fig.~\ref{fig:shapes_S5}. These examples represent a droplet structure ($\rho$
= 0.027 fm$^{-3}$, top panel), a rod-structure with a reduced density at $z=R$ as
compared to $z=0$ ($\rho$ = 0.0565 fm$^{-3}$, middle panel) and a rod-structure
with constant densities along the $z$-axis ($\rho$ = 0.0625 fm$^{-3}$, bottom panel) 

Comparing the DDRMF results of Fig.~\ref{fig:shapesDDRMF} with these density
distributions obtained from non--relativistic Skyrme calculations, 
we see that in case of Skyrme Hartree--Fock calculations
the different structures are more pronounced. Also we observe a slab structure in
the Skyrme Hartree--Fock approach, which is absent in the DDRMF calculations. This
differences may originate from the finite range of the NN interaction in the DDRMF
approach as compared to the zero-range forces of the Skyrme model.

In both cases we find  that the microscopic calculations in a Cartesian WS cell
for densities in the range of 0.01 fm$^{-3}$ to 0.1 fm$^{-3}$ lead to quite a
variety of shapes and quasi-nuclear structures with smooth transitions in between,
which may be characterized as quasi--nuclei, rod structures, slab structures, 
which are all embedded in a sea of neutrons and,  finally, the homogeneous
matter.  In the left part of table \ref{tab:shape_trans} the densities, at which 
the transitions from one shape to the other occur at $T=0$, are listed for Skyrme
HF calculations as well as for   the density dependent relativistic mean field
approach.

\begin{table}
\begin{center}
\begin{tabular}{|c|cc|cc|cc|cc|}
\hline
& \multicolumn{2}{c|}{Skyrme} 
  & \multicolumn{2}{c|}{DDRMF} &\multicolumn{2}{c|}{Skyrme} 
  & \multicolumn{2}{c|}{DDRMF}\\
\quad             	& HF   	 & TF      & H      &  TF & HF   & TF    & H    &  TF\\
\hline
droplet--rod   		& 0.042  & 0.066   & 0.057  & 0.048 & 0.040  & 0.048   & 0.044  & 0.045\\
rod--slab		& 0.070  & 0.078   & -      & -   & -      & -   	   & -      & -      \\
slab--homogeneous	& 0.080  & 0.085   & 0.064  & 0.061 & 0.065  & 0.048   & 0.044  & 0.045\\
\hline
&\multicolumn{4}{c|}{T = 0} &\multicolumn{4}{c|}{T = 5 MeV}\\
\hline
\end{tabular}
\end{center}
\caption{\label{tab:shape_trans}
          Comparison of densities at which shape transitions occur using the 
          Skyrme and the density dependent relativistic mean field (DDRMF) approach. Results are compared, 
	  employing the microscopic Hartree--Fock (HF), the mean-field or
	  Hartree (H) and the Thomas--Fermi (TF) approximation. Results displayed
	  in the left part of the table refer to a temperature of $T=0$, whereas
	  those listed in the right part were obtained for $T=5$ MeV.
          All entries are presented in fm$^{-3}$.}
\end{table}


Performing DDRMF calculations at the finite temperature $T=5$ MeV
we obtain for neutrons and protons density distributions as
displayed in the right column in Fig.~\ref{fig:shapesDDRMF}. Inspecting these
plots, it is clear that the geometrical structures observed for $T=0$ persist 
to some extent also at finite temperatures as large as $T=5$ MeV. It is also
obvious, however, that finite temperature effects tend to dissolve these
structures: The structures in the right column at corresponding or even lower
densities are less pronounced than those in the left column of 
Fig.~\ref{fig:shapesDDRMF}. The same features can also be observed for the Skyrme
Hartree--Fock approach, as can be seen from Fig.~\ref{fig:shapes_S5}.

The transition densities for all calculations in 
the WS cell at a temperature of
$T=5$ MeV are summarized in the right part table \ref{tab:shape_trans}.
If we compare the Skyrme HF transition densities at $T=5$ MeV 
with the zero temperature results,
the transition densities are
lower at finite temperature and the slab structure disappears.
The results obtained from the DDRMF approach 
do not any more show a rod structure, 
but the transition from the droplet structure 
to homogeneous matter occurs via
a grid structure, which can be regarded as transition
to the homogeneous phase like in the zero temperature case.


\begin{figure}
\begin{center}
  \mbox{  \includegraphics[width = 8cm]{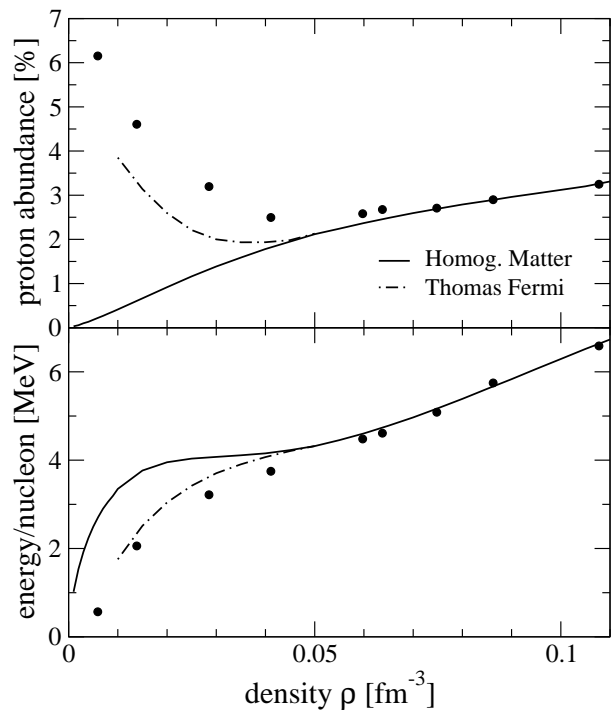}  }
\end{center}
\caption{\label{fig:E-DDRMF_T0}
Proton--abundances and energy per nucleon resulting from
density dependent relativistic 
mean-field calculations (DDRMF) at different densities. 
The results evaluated in cubic Wigner Seitz cells (circles) 
are compared to those of homogeneous infinite matter (solid lines)
and corresponding Thomas--Fermi calculations (dashed lines). }
\end{figure}

A comparison of the proton abundances and the energy per nucleon 
obtained within the DDRMF model in a Wigner--Seitz cell for zero temperature
are displayed in Fig.~\ref{fig:E-DDRMF_T0}.
The solid lines represents the homogeneous matter results, while the
circles display the results obtained from the microscopic
density dependent relativistic mean field calculation in the cubic box.
At densities larger than about 0.08 fm$^{-3}$ the proton abundances and energies
coincide for homogeneous matter and the microscopic calculation, which implies
that for densities larger than 0.08 fm$^{-3}$ the formation of inhomogeneous
structures does not provide any gain in energy for baryonic matter in
$\beta$-equilibrium. For lower densities, however, the non--homogeneous
structures provide a gain in binding energy up to 1.7 MeV per nucleon. 
These quasi-nuclear structures, embedded in a sea of neutron matter, are also
responsible for the enhancement of proton abundances at small global densities:
the protons are constrained to the regions of enhanced densities.

One can try to simulate these features resulting from the non--homogeneous
structures using the Thomas--Fermi approximation. Like in a
Ref.~\cite{Goegelein:2007}  we want to compare the DDRMF results with the
Thomas--Fermi calculation based on a local density approximation  of the energy
density for homogeneous nuclear matter. In our Thomas--Fermi (TF) calculations we use
a parameterization for the density distributions like in \cite{Oyamatsu:2007} 
\begin{equation}
\rho_i(r) = 
    \begin{cases}
	(\rho_i^{in} - \rho_i^{out} )
	\left[ 1 - \left( \frac{r}{R_i} \right) ^{t_i} \right]^3 + \rho_i^{out}, & r < R_i \\
	\rho_i^{out}, & R_i \leq r,\ \ \ \ i = p,n,
    \end{cases}    
\label{eq:param1}
\end{equation}
where the central density $\rho_i^{in}$, the peripheral density $\rho_i^{out}$, 
the structure radius $R_i$ and an exponent $t_i$ are the variational parameters
completed by the box-size $R$ and the proton abundance.
As an alternative a Wood--Saxon type density parameterization has been considered, 
but it turned out that the results are comparable and only the 
surface energy constant $F_0$ has to be readjusted.
For the description of rod--shape quasi--nuclear structures cylindrical
coordinates are used to parameterize the dependence of the densities on the radial
coordinate in a way corresponding to eq.(\ref{eq:param1}).
In the case of quasi--nuclear structures in form of slab--shapes these
parameterizations are considered for the dependence of the densities on the
Cartesian $z$-coordinate.

Assuming those density distributions, the TF energy is calculated as a sum of
the bulk energy, i.e. the integrated nuclear--matter energy densities, 
the Coulomb energy, plus the
contribution of a surface term of the form \cite{Oyamatsu:2007,Oyamatsu:1993}
\begin{equation}
E_{\text{surf}} = F_0\,\int_{\text{WS-cell}} d^3r\,\left|
\boldsymbol{\nabla}\rho\right|^2 \,.
\label{eq:esurf}
\end{equation}
The parameters of the density distribution in(\ref{eq:param1})
are varied to minimize the energy of the system under
consideration. 
The Parameter $F_0$ for the surface energy term in
(\ref{eq:esurf}) has been adjusted in such a way that the TF calculation reproduces 
the experimental binding energy of the nucleus $^{208}Pb$, which has also been
reproduced in the DDRMF calculation.  
This adjustment leads to a value for $F_0$ of 61.0 MeV fm$^{5}$, which is rather
similar to the value of  68.3 MeV fm$^{5}$, derived in \cite{Goegelein:2007} 
for Skyrme HF case. 

Such TF calculations lead to non--homogeneous structures similar to those
calculated in DDRMF and we obtain transitions between the various shapes at the
densities, which are listed in table~\ref{tab:shape_trans}.
Comparing the Thomas--Fermi results to the one obtained from 
the microscopic DDRMF calculations in Fig.~\ref{fig:E-DDRMF_T0}
in the density range below 0.05 fm$^{-3}$
it is observed 
that the proton abundance from the Thomas--Fermi calculations
are lower and the energy is larger as obtained from the corresponding
microscopic DDRMF results.
The surface energy constant $F_0$ may be readjusted to reproduce the
microscopic DDRMF results for the nuclear structures in $\beta$--equilibrium.
This leads to a value of $F_0$, which is only about half the value for $F_0$
derived from the fit to the properties of $^{208}$Pb. This suggests to consider
a surface constant $F_0$, which depends on the isospin asymmetry of the system.
This feature is very similar to the one observed 
applying the Skyrme Hartree--Fock approach\cite{Goegelein:2007}.


\begin{figure}
\begin{center}
\mbox{
\includegraphics[width =8cm]{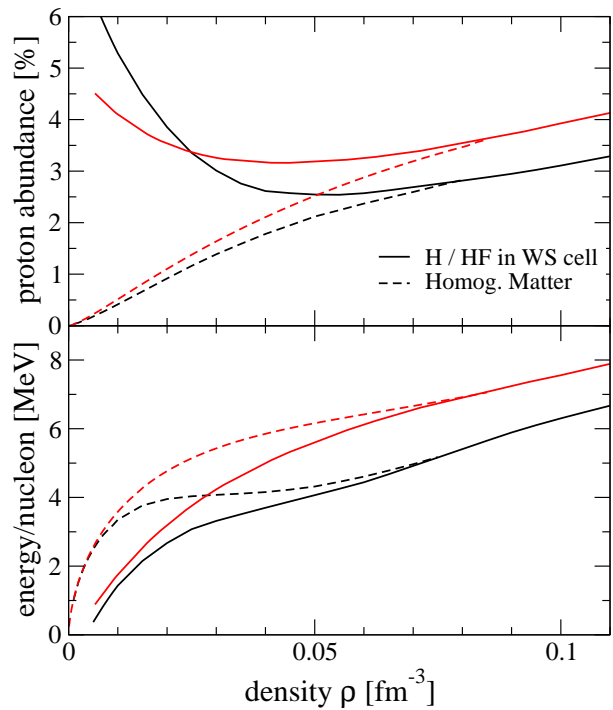} 
}
\end{center}
\caption{\label{fig:RSEcomp} (Color online) 
Comparison of proton abundances and energy per nucleon obtained from
DDRMF (black color) and Skyrme Hartree--Fock calculations (red color).
Solid lines are obtained from a polynomial fit to 
 the DDRMF or the Skyrme Hartree--Fock calculation
in the cubic Wigner--Seitz cell, while dashed lines represent the
results for homogeneous infinite nuclear matter.}
\end{figure} 

To enable further explorations on the differences between the predictions for the
``pasta-phase'' derived from DDRMF and Skyrme HF calculations, results of those
two approaches for the energy per nucleon and the proton abundances are directly
compared in  
Fig~\ref{fig:RSEcomp} 
The results of both models show a similar overall behavior and both predict a
transition to the homogeneous phase at a density of about 0.08 fm$^{-3}$.
Looking into details, however, one finds rather interesting differences: The
energy per nucleon is larger for the Skyrme calculation at all densities
considered in Fig~\ref{fig:RSEcomp}. This is in line with the fact that the
Skyrme force SLy4, which we have used here, yields a parameter for the
symmetry energy $a_S$ of 32 MeV\cite{chabanat}, which is larger than the one
derived from DDRMF (see table~\ref{tab:NM_prop_vD_param}). Note, however, that
the symmetry energy coefficient is determined at saturation density, whereas
here we are comparing structures at half the saturation density and below. A 
larger symmetry energy at those small density would also explain the larger
proton abundances obtained in the Skyrme approach.  

The energy gain due
to the formation of non--homogeneous structures is slightly larger for the Skyrme
approach at medium
densities of 0.02 to 0.08 fm$^{-3}$. This is in line with our observation made
in the discussion above, that the geometrical structures determined in Skyrme HF
are typically more pronounced than those extracted from DDRMF. At very small
densities the DDRMF leads to a larger gain in energy forming non--homogeneous
matter distributions. At these densities the DDRMF calculations in the WS cell
also predict a larger proton abundance than the Skyrme approach.
 
The parameters of the Skyrme interaction have only been adjusted to the data of
finite nuclei and the saturation point of symmetric matter, i.e.~data of nuclear
systems at normal densities and proton-neutron asymmetries. The parameterization
of the DDRMF on the other side also reproduces those data but is furthermore
based on a realistic NN interaction. Therefore, we trust in the predictions of the
DDRMF approach for this region of small densities and large proton-neutron
abundances more than in those based on the Skyrme Hamiltonian.
   
\begin{figure}
\begin{center}
  \mbox{  \includegraphics[width = 8cm]{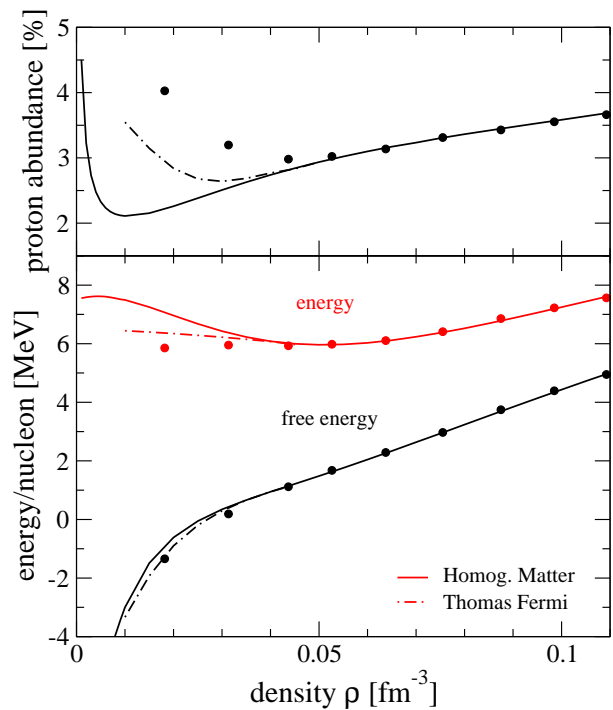}  }
\end{center}
\caption{\label{fig:E-DDRMF_T5}
Proton abundances and energy per nucleon resulting from
density dependent relativistic
mean field calculations at the temperature $T=5$ MeV. 
The results evaluated in cubic Wigner--Seitz cells (circles) 
are compared to those of homogeneous infinite matter
(solid lines). }
\end{figure}

Finally, we are going to discuss the effects of finite temperature. For that
purpose Fig.~\ref{fig:E-DDRMF_T5} displays results for proton abundances and the
energy per nucleon evaluated in the DDRMF approach at a temperature of $T=5$
MeV. 
The lower panel in Fig.~\ref{fig:E-DDRMF_T5} shows two curves: 
the energy per nucleon (upper curve)
and the free energy per nucleon (lower curve).
Comparing with the results obtained from zero temperature DDRMF calculations
we recognize that the results for $T=5$ 
show a minimum for the energy per nucleon 
at a density of about 0.05 fm$^{-3}$, while
the zero temperature results show no minimum.
However, the free energy, which is the energy of consideration,
is lower than the energy in the $T=0$ case and rises with
a larger slope, which means that the finite temperature increases the pressure
derived from the free energy.


The energies resulting from the DDRMF calculations, which allow for
non--homogeneous structure yield a gain in energy for densities below 0.05
fm$^{-3}$. This means that a temperature of 5 MeV lowers the critical density
for the formation of local structures from 0.08 fm$^{-3}$ at $T=0$ to 0.05 
fm$^{-3}$ at $T=5$ MeV. At values below this critical density, the gain in the
free energy at finite temperature turns out to be considerably smaller than in 
the zero temperature limit. This reflects the feature that finite temperature
effects tend to suppress the formation of density fluctuations. Note, that these
small differences between the free energy of the homogeneous and inhomogeneous
structures is to quite some extent due to the enhancement of the entropy in the
inhomogeneous case. We observe the  same features in Skyrme HF calculations
at finite temperature. Finite temperature also reduces the variety of
geometrical shapes in the Skyrme HF as well as in the DDRMF approach (see
table~\ref{tab:shape_trans}).   

Fig.~\ref{fig:E-DDRMF_T5} also displays results from Thomas-Fermi calculations.
In this case the temperature dependence is contained only in the
the bulk contribution, i.e. the integrated nuclear--matter energy densities
calculated at the finite temperature. The parameter for the surface energy $F_0$
is still adjusted to reproduce the properties of $^{208}$Pb at $T=0$. As
discussed above one could now try to fit this constant $F_0$ in such a way that
we reproduce results for the energy, the free energy and proton abundances of
the DDRMF approach. 
This readjustment leads to a value of 
27 MeV fm$^5$ for the parameterization of eq.(\ref{eq:param1}).
The value is lower by about 15\% compared to the corresponding 
one at zero temperature. 
Hence a slight reduction of the surface energy 
constant $F_0$ could improve the Thomas--Fermi
approach at finite temperature.

\section{Conclusion}

Density dependent relativistic mean field (DDRMF) calculations have been
performed to study the structure of baryonic matter in $\beta$-equilibrium with
electrons in a region of densities between 0.01 and 0.1 nucleons  fm$^{-3}$. A
parameterization of the density dependent meson-nucleon coupling constants
has been developed, which is based on microscopic Dirac Brueckner Hartree Fock
(DBHF) calculations and reproduces the saturation point of nuclear matter as
well as the bulk properties of finite nuclei. Since this parameterization yields
a good description of normal nuclei, i.e. baryonic matter at normal density and
small proton - neutron asymmetries but is also based on a realistic interaction
which describes the scattering of two nucleons in the vacuum, it should give
more reliable results for the baryonic structure at small densities and large
proton-neutron asymmetries than corresponding calculations, which are based on
purely phenomenological Hamiltonians like the Skyrme forces. The 
parameterization of the coupling constants in terms of analytic functions makes
it easily accessible.

The DDRMF calculations are performed for zero temperature as well as finite
temperature in a periodic lattice of Wigner-Seitz (WS) cells of cubic shape.
Pairing correlations are taken into account within the framework of the finite
temperature BCS approximation assuming a density-dependent pairing force of zero
range.

At densities below 0.08 nucleons fm$^{-3}$ the DDRMF approach predicts
non-homogeneous structures, which are similar to those predicted by Thomas-Fermi 
or Skyrme Hartree-Fock calculations for the ``pasta-phase'' of matter in
the crust of neutron stars. The structures obtained in the DDRMF calculations,
however, are typically less pronounced than the corresponding structures
resulting from Skyrme Hartree--Fock calculations. This may be a consequence of
the finite range of the interaction in DDRMF as compared to the zero-range
Skyrme forces. Also it turns out that the gain in energy due to the formation of
non--homogeneous density distributions is slightly smaller in the DDRMF as
compared to Skyrme HF.
Also the DDRMF yields a smaller symmetry energy at the low densities, which
provides smaller proton abundances.

The effects of finite temperature tend to reduce the disposition of the baryonic
matter to form non--homogeneous structures. This is driven by the larger entropy
of the non--homogeneous matter as compared to the homogeneous phase. This leads
to a reduction of the critical density for the formation of non--homogeneous
structures. It drops from 0.08 nucleons fm$^{-3}$ at $T=0$ to 0.05 nucleons 
fm$^{-3}$ at $T=5$ MeV. At densities below the critical densities, the gain in
the free energy of the non--homogeneous as compared to the homogeneous 
realization is smaller and the resulting geometrical structures are less
pronounced in the case of finite temperature. 

An attempt has been made to reproduce the results of the microscopic Skyrme HF
and DDRMF calculation by means of the Thomas-Fermi (TF) approximation. Main 
features like the critical densities for the formation of geometrical structures
of various shapes, the energies and the proton abundances can be reproduced, if
one considers a parameter for the surface term, which is reduced with increasing
proton asymmetry and with temperature.   

This work has been supported by the European Graduate School ``Hadrons in Vacuum in
Nuclei and Stars'' (Basel, Graz, T\"{u}bingen), which obtains financial support by
the DFG and received support from the GSI project TUEMUE and by the Spanish Ministry of
Education and Science under grant no. SB-2005-0131.



\begin{thebibliography}{99}
\bibitem{bal1} M. Baldo, {\it Nuclear Methods and the Nuclear Equation of State,}
Int. Rev. of Nucl. Phys, Vol. 9 (World Scientific, Singapore, 1999).

\bibitem{her1} H. M\"uther and A. Polls, Prog. Part. Nucl. Phys.
{\bf 45}, 243 (2000).

\bibitem{pieper02} S.C. Pieper, K. Varga, and R.B. Wiringa, Phys. Rev. C
\textbf{66}, 044310 (2002).  

\bibitem{anast} M.R.\ Anastasio, L.S.\ Celenza, W.S.\ Pong, and C.M.\ Shakin,
Phys. Rep. \textbf{100}, {327} (1983).

\bibitem{brock} R. Brockmann and R. Machleidt, Phys. Rev. C \textbf{42}, {1965}
(1990).

\bibitem{malf1} B.\ Ter Haar and R.\ Malfliet, Phys. Rep. \textbf{149}, 
{207} (1987).

\bibitem{weigel} H.\ Huber, F.\ Weber, and M.K.\ Weigel, Phys. Lett. B 
\textbf{317}, {485} (1993).

\bibitem{serot} B.D.\ Serot and J.D.\ Walecka, Adv.\ Nucl.\ Phys.
\textbf{16}, {1} (1986).

\bibitem{plohl06}
  O.~Plohl and C.~Fuchs,
  Phys.\ Rev.\  C {\bf 74}, 034325 (2006).


\bibitem{Li:2006gr}
  Z.~H.~Li, U.~Lombardo, H.~J.~Schulze, W.~Zuo, L.~W.~Chen and H.~R.~Ma,
  Phys.\ Rev.\  C {\bf 74}, 047304 (2006).

\bibitem{sk1} T.H.R. Skyrme, Nucl. Phys. {\bf 9}, 615 (1959).

\bibitem{sk2} D. Vautherin and D.M. Brink, Phys. Rev. C{\bf 5}, 626, (1972).

\bibitem{bv81} P. Bonche and D. Vautherin, Nucl Phys. A{\bf 372}, 496 (1981).

\bibitem{Montani:2004}{F. Montani, C. May, and H. M\"{u}ther, Phys. 
Rev. C {\bf 69}, 065801 (2004).}

\bibitem{Goegelein:2007}{P. G\"{o}gelein and H. M\"{u}ther, Phys. Rev. (2007) in
print.}

\bibitem{Schiller:2001}{E. Schiller and H. M\"{u}ther, Eur. Phys. J. {\bf A11}, 15 (2001).}

\bibitem{Hofmann:2001}{F. Hofmann, C.M. Keil, and H. Lenske, Phys. Rev. C \textbf{64}, 034314  (2001).}

\bibitem{vanDalen:2007}{E.N.E. van Dalen, C. Fuchs, and  A. Faessler, Eur. Phys. J. A \textbf{31}, 29 (2007).}


\bibitem{Klaehn:2006}{T. Kl\"ahn, D. Blaschke, S. Typel, E.N.E. van Dalen, A. Faessler, C. Fuchs, T. Gaitanos, H. Grigorian, A. Ho, E.E. Kolomeitsev, M. C. Miller, G. R\"opke, J. Tr\"umper,D. N. Voskresensky, F. Weber, and H. H. Wolter, Phys. Rev. C \textbf{74}, 035802 (2006).}

\bibitem{Bjorken_Drell:1964}{J.D. Bjorken and S.D. Drell, Relativistic Quantum Mechanics, McGraw-Hill, New York, 1964.}

\bibitem{Fuchs:1995}{C. Fuchs, H. Lenske, and H.H. Wolter, Phys. Rev. C \textbf{52}, 3043 (1995).}

\bibitem{Fritz:1994} R. Fritz and H. M\"uther, Phys. Rev. C \textbf{49}, 633
(1994);{R. Fritz, Ph-D thesis, T\"{u}bingen (1994).}

\bibitem{Goodman:1981}{A.L. Goodman, Nucl. Phys. {\bf A352}, 30 (1981).}

\bibitem{Shen:1998a}{H. Shen, H. Toki, K. Oyamatsu, K. Sumiyoshi, Nucl. Phys. \textbf{A637}, 435 (1998).}

\bibitem{Shen:1998b}{H. Shen, H. Toki, K. Oyamatsu, K. Sumiyoshi, Prog. Theor. Phys. \textbf{100}, 1013 (1998).}


\bibitem{Dejong:1998}{F. de Jong and H. Lenske, Phys. Rev. C \textbf{58}, 890 (1998).}


\bibitem{vanDalen:2004b}{E.N.E. van Dalen, C. Fuchs, and  A. Faessler, Nucl. Phys. \textbf{A744}, 227 (2004).}

\bibitem{Gross:1999}{T. Gross-Boelting, C. Fuchs, and Amand Faessler, Nucl. Phys. \textbf{A648}, 105  (1999).}

\bibitem{Ring:1996}{P. Ring, Prog. Part. Nucl. Phys. \textbf{37}, (1996) 193.}

\bibitem{Sturm:2001}{C. Sturm {\it et al.} [KaoS Coll.],
Phys. Rev. Lett. \textbf{86}, 39 (2001); C. Fuchs, A. Faessler, E. Zabrodin, and Y.M. Zheng,
Phys. Rev. Lett. \textbf{86}, 1974 (2001); C. Fuchs, Prog. Part. Nucl. Phys. \textbf{56}, 1 (2006).}

\bibitem{Khoa:2005ze}
 D.~T.~Khoa, W.~von Oertzen, H.~G.~Bohlen and S.~Ohkubo,
  J.\ Phys.\ G {\bf 33}, R111 (2007);
  D.~T.~Khoa {\it et al.},
  Nucl.\ Phys.\  A {\bf 759}, 3 (2005)

\bibitem{Oyamatsu:2007}{K. Oyamatsu and K. Iida, Phys. Rev. {\bf C75}, 015801 (2007).}

\bibitem{Oyamatsu:1993}{K. Oyamatsu, Nucl. Phys. {\bf A561}, 431 (1993).}

\bibitem{chabanat} E. Chabanat, P. Bonche, P. Haensel, J. Meyer, and R.
Schaeffer, Nucl. Phys. {\bf A 635}, 231 (1998). 
\end{thebibliography}
\end{document}